# AC-Tolerant Multifilament Coated Conductors

G. A. Levin, P. N. Barnes, and N. Amemiya

*Abstract*—We report the magnetization losses in an experimental multifilament coated conductor. A 4 mm wide and 10 cm long YBCO coated conductor was subdivided into eight 0.5 mm wide filaments by laser ablation and subjected to post-ablation treatment. As the result, the hysteresis loss was reduced, as expected, in proportion to the width of the filaments. However, the coupling loss was reduced dramatically, and became practically negligible, in the range of a sweep rate up to 20 T/s. This represents a drastic improvement on previous multifilament conductors in which often the coupling losses became equal to the hysteresis loss at a sweep rate as low as 3-4 T/s. These results demonstrate that there is an effective and practical way to suppress coupling losses in coated multifilament conductors.

*Index Terms*— ac loss, coated conductors, coupling loss.

## I. INTRODUCTION

In recent years the first steps have been taken to make and test ac-tolerant multifilamentary YBa$_2$Cu$_3$O$_{6+x}$ (YBCO)-coated conductors with the goal in mind of introducing them into future ac power applications. See, for example, the partial listing of references given for this subject, Refs. [1]-[10]. A way to reduce ac losses is to divide the superconducting layer into a large number of filaments (stripes) segregated by non-superconducting resistive barriers [11]. This seemingly simple solution has several pitfalls, some already revealed in experiments, others are anticipated, and yet a few more may appear later.

In the first multifilamentary samples produced by striation of conventional (uniform) YBCO-coated conductors presently manufactured by industry, the loss in an applied time-varying magnetic field is reduced as expected. However, it is only at a relatively low sweep rate, B$f$. Here B is the amplitude of the applied field and $f$ is its frequency. The culprit can be readily identified as the coupling loss. The relatively low resistance of the grooves segregating the superconducting filaments allows current to flow between these stripes and dissipate energy.

An overriding consideration in choosing the means to overcome the emerging problems with coated conductors is its economic viability. The manufacturers of conventional coated conductors are striving to approach the threshold at which these conductors may become competitive with copper wires. Therefore, the additional processes by which we expect to make striated conductors must not impose too great of an economic penalty on the currently existing technology for coated conductors.

Here we describe an approach for combating the coupling loss which appears promising in its effectiveness and practicality. In essence, it can be described simply as a deliberately induced and accelerated corrosion of the metal in the grooves which segregate the superconducting filaments. Since most metal oxides have high resistivity, they can then drastically reduce the coupling current and the attendant energy dissipation.

## II. PRINCIPLE OF COUPLING LOSS REDUCTION

An up-to-date description of coupling loss is given by Carr in Ref.[12]. The coupling component of power loss per unit length can be expressed phenomenologically as

$$Q_c = \frac{\pi^2}{6} \frac{(BfL)^2}{R_{eff}} W \:. \qquad (1)$$

Here $L$ is either the length of a straight sample or half of the twist pitch, W is the width of the conductor and $R_{eff}$ is the phenomenological coupling resistance. Equation (1) can be obtained either as a back-of-envelope estimate [5] or by a more refined derivation as in [12]. Here, we keep the numerical coefficient for consistency.

Given the mechanical properties of a wide thin metal tape and the brittleness of a thin YBCO film, it is highly desirable for any potential application to have the freedom of making the twist pitch length as large as necessary, in accordance with the requirements of the application. The reduction of coupling loss by shortening the twist pitch length should not be the dominant consideration in designing coils and other deveices. Therefore, increasing the coupling resistance is the primary mechanism by which one would try to suppress the coupling loss.

For multifilamentary conductors produced by laser ablation [3],[5] the reported value of $R_{eff} \approx \rho/d$, where ρ and $d$ are the resistivity and thickness of the substrate respectively. At first glance it may appear that in order to increase the coupling resistance, the resistivity of the bulk of the substrate would have to drastically increase. Fortunately, this is not so. Unlike the eddy currents, that fill more or less uniformly the whole volume of a metal object exposed to time-varying magnetic field, the coupling currents originate at the edges of the superconducting stripes [13]. Therefore, if these edges are insulated, the coupling currents will be suppressed, even though the bulk of the substrate may still have a low resistivity.

Figure 1(a) shows a microphotograph of a profile of a groove made by laser ablation in a coated conductor sample with Hastelloy substrate manufactured by SuperPower Inc. [14]. The parameters of laser ablation are described in [4],[5],[15].

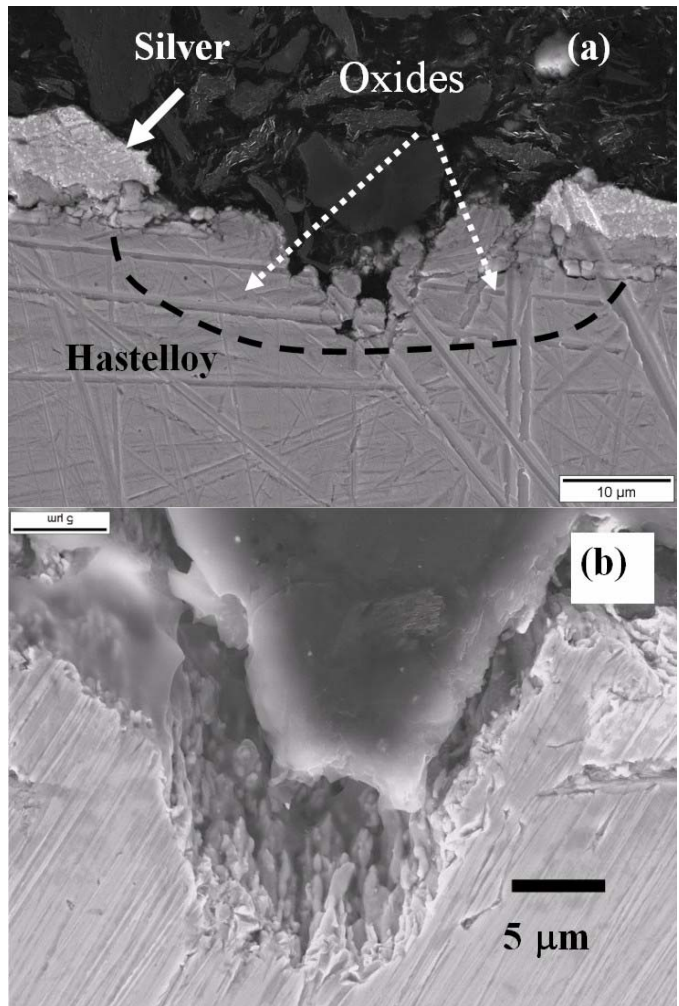

*Fig. 1 (a) Profile of a groove made by laser ablation. The Hastelloy substrate and silver cap layer are indicated. The width of the groove is about 30 μm. The dashed line indicates schematically the volume of metal that needs to be oxidized in order to insulate the edges of YBCO stripes. (b) A close up view of another groove. Multiple conical spikes (penitentes) are clearly visible.*

Laser ablation carried out in the regimes described in [4],[15] creates irregular shaped grooves in the substrate, completely removing silver and YBCO from the exposed area. The width of the groove is about 30 microns. Figure 1(b) shows a close-up view of the surface of a groove. Clearly seen are multiple cone-shaped protrusions, also known as penitentes. These are universal features of both laser- and photo-ablation [16].

The highly disrupted and rough surface of the grooves creates favorable conditions for corrosion (oxidation). Oxygen (or other corrosive agents in gaseous or liquid form) can penetrate deeper into the substrate through the microscopic cracks in the large exposed surface area. Oxides can then nucleate at favorable sites, such as dislocations and impurity atoms. Hastelloy is a corrosion resistant alloy. Nevertheless, the damage to the surface evident in Figs. 1(a,b) render it vulnerable to accelerated corrosion to a much greater extent then it would have been under normal circumstances, i.e. an undamaged surface.

The superconducting film is deposited on an insulating buffer. Therefore, only a relatively small volume of the groove has to be corroded in order to seal the superconducting stripe at its edges, as shown schematically in Fig. 1(a). Hastelloy is a Ni-Cr-Mo-W alloy and its corrosion leads to formation of high resistivity non-metallic oxides such as $NiO$, $Cr_2O_3$, etc. or compound oxides such as spinels $NiO \cdot Cr_2O_3$ that form at higher temperatures.

### III. EXPERIMENTAL RESULTS

The samples were made from a 4 mm wide uniform coated conductor provided by SuperPower Inc. [14]. Striation was accomplished by laser micromachining using a frequency tripled diode-pumped solid-state Nd:$YVO_4$ laser at a 355 nm wavelength [4],[5],[15]. The 10 cm long, 4 mm wide conductor was divided into 8 stripes, so that the width of an individual superconducting filament was close to 0.5 mm. For comparison we also measured losses in a non-striated (control) sample of the same width and length. Both samples were initially parts of a longer piece of tape.

After ablation the striated sample was annealed in flowing oxygen in order to induce oxidation in the exposed grooves. The parameters of annealing were similar to those required to oxygenate YBCO and maximize its critical temperature. The temperature of the sample was raised gradually to $550^0$C over 2-3 hours, kept at 500-$550^0$C for about 2-3 hours, and reduced to room temperature over another 3 hours.

The losses in the striated as well as the non-striated (control) samples were measured by placing them inside the bore of a dipole magnet that generated a time-varying magnetic field as described in Ref. [2]. The transport critical current was measured in both conductors using the conventional 1μV/cm criterion. The self-field critical current of the control sample was 61 A. Before striation the critical current in what will become the 8-filament sample was 55 A. After striation and oxygen annealing its critical current was reduced by 10% to 49 A.

Figure 2 presents the data for both samples in the form of energy loss per cycle, per meter length. Since coupling loss has quadratic frequency dependence, its presence in any appreciable amount would have resulted in a very noticeable frequency dependence of the loss per cycle. However, the frequency dependence is quite weak, indicating that in both samples only hysteresis loss is present. This is in large contrast to the previous results obtained on similar samples, Refs. [3]-[5], that were not subjected to post-ablation oxygenation. In fact, the frequency dependence of $Q/f$ is greater in the control sample which may be attributable to eddy current losses in the silver cap layer. In the 8-filament sample the cap layer is striated as well and the eddy current loss is suppressed.

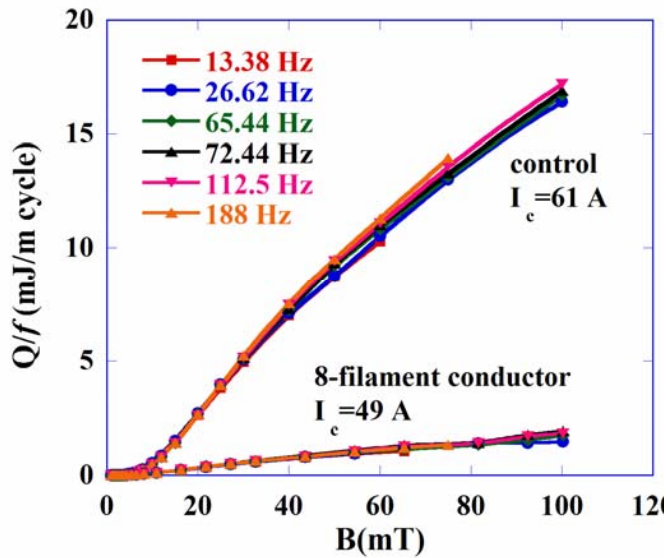

*Fig. 2. Energy loss per meter length, per cycle vs amplitude of the applied field in the control and in the 8-filament samples at different frequencies of applied magnetic field.*

An objective criterion of effectiveness of striation is the amount of power loss per unit of critical current $Q/I_c$. The reduction of dissipated energy by itself does not make it clear whether this is the result of striation or simply the result of degradation of the superconducting film. When we normalize the data in Fig. 2 by the respective critical current, the specific power loss per unit of critical current is almost eight times smaller in the 8-filament conductor than in the control one[10]. The maximum possible reduction of the specific loss with respect to non-striated conductor is by a factor 1/N, where N is the number of stripes. This indicates that all stripes in our conductor can carry their share of current and no unnecessary damage was inflicted upon them by striation. Further confirmation of the effectiveness of post-ablation oxidation was obtained by Demencik *et al.* [17] on a different set of samples.

### IV. FURTHER DEVELOPMENT

The results presented here illustrate the principle of coupling loss reduction method based on induced accelerated corrosion of the grooves. However, these samples cannot be considered truly conductors because they are not stabilized. Copper stabilizer has to be striated in the same pattern as the underlying superconducting film. An obvious, although not necessarily the most efficient, way to accomplish that is to apply laser micromachining to a fully stabilized uniform coated conductor and to cut grooves through the copper, YBCO, and buffer. Figure 3 shows several grooves cut in a copper sheet by laser ablation using a different laser system than the one described in [4], [5], and [15]. The grooves are much cleaner than that in Fig. 1 with almost vertical walls. The width of the grooves is about the same as in Fig. 1a (30 μm).

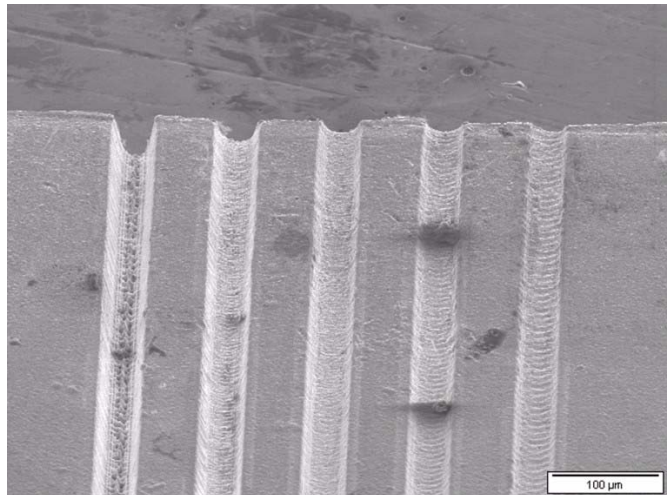

*Fig. 3. Grooves cut in copper stabilizer of a coated conductor by laser ablation at different power levels.*

If this method of striation is adopted, the corrosion of the substrate at the bottom of the grooves may have to be carried out differently. The prolonged exposure to oxygen at high temperature of the stabilized conductor may be undesirable. Formation of a relatively small volume of highly resistive compounds on the walls and at the bottom of a groove (passivation of the groove) can be facilitated by exposure of the conductor to an aggressive corrosive agent such as vapors of nitric or sulfuric acid for example. The surface of the stabilizer over the current-carrying stripes can be easily protected from such exposure.

Another possibility to stabilize a striated coated conductor is closely related to the oxidation of the grooves described above. If the stabilizer is applied to the conductor by electroplating [14], the non-conducting surface of the grooves will prevent growth of the copper layer over them. It is therefore seems possible to incorporate post ablation oxidation into the currently developed technological chain with only relatively small modifications in order to manufacture fully-stabilized multifilamentary coated conductors.

### V. SUMMARY

We have presented a first demonstration of a method to suppress the coupling losses in multifilamentary coated conductors by oxidation of the grooves segregating the superconducting filaments. Corrosion leads to formation of non-conducting compounds on the walls and at the bottom of the grooves that insulate the edges of the superconducting filaments from the substrate and from each other. This method is very effective in accomplishing its purpose. Coupling loss has been greatly suppressed in a 10 cm long conductor to a negligible level at a sweep rate B$f$ as high as 20 T/s. This method (or its modifications) with the use of corrosive agents other than oxygen, and at lower temperatures, has a potential to be compatible with the current technology of manufacturing of coated conductors.

ACKNOWLEDGMENT

We thank J. W. Kell, N. Yust, and S. Sathiraju for help with microphotography of the samples.

REFERENCES

[1] C.B. Cobb, P.N. Barnes, T.J. Haugan, J. Tolliver, E. Lee, M. Sumption, E. Collings, C.E. Oberly, Physica C **382**, 52 (2002).
[2] N. Amemiya, S. Kasai, K.Yoda, Z. Jiang, G. A. Levin, P. N. Barnes, and C. E. Oberly, Supercond. Sci. Technol. 17, 1464 (2004).
[3] G. A. Levin, P. N. Barnes, N. Amemiya, S. Kasai, K.Yoda, and Z. Jiang, Appl. Phys. Lett. 86, 072509 (2005).
[4] M.D. Sumption, E.W. Collings, and P.N. Barnes, Supercond Sci. Technol., **18**, 122-134 (2005).
[5] G. A. Levin, P. N. Barnes, N. Amemiya, S. Kasai, K. Yoda, Z. Jiang, and A. Polyanskii, J. Appl. Phys. **98**, 113909 (2005) and references therein.
[6] M. Majoros, B. A. Glowacki, A. M. Campbell, G. A. Levin, P. N. Barnes, and M. Polak, IEEE Trans. Appl. Supercond. **15**, 2819 (2005).
[7] W.J. Carr, and C.E. Oberly, IEEE Trans. on Appl. Supercond. **9**, 1475 (1999).
[8] M. Polak, E. Usak, L. Jansak, E. Demencik, G. A. Levin, P. N. Barnes, D. Wehler, and B. Moenter, e-print cond-mat/0602422.
[9] S. P. Ashworth and F. Grilli, Supercond Sci. Technol., **19**, 227-232 (2006).
[10] G. A. Levin, P. N. Barnes, N. Amemiya, Z. Jiang, K. Yoda, and F. Kimura, "Multifilament YBa$_2$Cu$_3$O$_{6+x}$-coated conductors with minimized coupling losses", .Appl. Phys. Lett. 89, 012506 (2006).
[11] W.J. Carr, and C.E. Oberly, IEEE Trans. on Appl. Supercond. **9**, 1475 (1999).
[12] W.J. Carr, "Loss in striated coated conductor", to be published
[13] W.J. Carr, and C.E. Oberly, Supercond. Sci. Technol. **19,** 64 (2006).
[14] Y.-Y. Xie, A. Knoll, Y. Chen, Y. Li, X. Xiong, Y. Qiao, P. Hou, J. Reeves, T. Salagaj, K. Lenseth, L. Civale, B. Maiorov, Y. Iwasa, V. Solovyov, M. Suenaga, N. Cheggour, C. Clickner, J.W. Ekin, C. Weber and V. Selvamanickam, *Physica C* **426-431,** 849-857 (2005).
[15] K. E. Hix, M. C. Rendina, J. L. Blackshire, and G. A. Levin, e-print cond-mat/0406311
[16] V. Bergeron, C. Berger,and M. D. Betterton, "Controlled Irradiative Formation of Penitentes" Phys. Rev. Lett. 96, 098502 (2006)
[17] E. Demencik, P. Usak, S. Takacs, I. Vavra, M. Polak, G. A. Levin, and P. N. Barnes, "Visualization of coupling current paths in striated YBCO coated conductors at frequencies up to 400 Hz", to be published.